\documentstyle[preprint,pre,aps]{revtex}
\newcounter{saveeqn}

\title{Dynamics of the maximum marginal likelihood hyper-parameter 
estimation in image restoration : \\
gradient descent vs. EM algorithm}

\author{Jun-ichi Inoue} 
\address{Complex Systems Engineering, Graduate School of 
Engineering, \\
Hokkaido University, N13-W8, Kita-ku, Sapporo 060-8628, Japan}
\author{Kazuyuki Tanaka} 
\address{Department of Computer and Mathematical Science, 
Graduate School of Information Science, \\
Tohoku University, Aramaki-aza-aoba 04, Aoba-ku, Sendai 
980-8579, Japan }

\date{\today}
\begin{document}
%%%%%%%%%%%%%%%%%%%%%%%%%%%%%%%%%%%%%%%%%%%%%%%%%%%%%%%%%%%%%
%\draft
\maketitle
\thispagestyle{empty}
\begin{abstract}
Dynamical properties of image restoration 
and hyper-parameter estimation are investigated 
by means of statistical mechanics.  
We introduce an exactly solvable model for image restoration and derive 
differential equations with respect to macroscopic quantities.  
From these equations, we evaluate relaxation processes 
of the system to the equilibrium state. 
Our statistical mechanical approach also enable us to 
investigate the hyper-parameter estimation by means of 
maximization of marginal likelihood by using gradient 
decent and EM algorithm 
from dynamical point of view.  
\end{abstract}
\mbox{}
\clearpage
\setcounter{page}{1} 
%%
%%%%%%%%%%%%%%%%%%%%%%%%%%%%%%%%%%%%%%%%%%%%%%%%%%%%%%%%%%%%%%%%%%%%
\section{Introduction}
%%%%%%%%%%%%%%%%%%%%%%%%%%%%%%%%%%%%%%%%%%%%%%%%%%%%%%%%%%%%%%%%%%%%
%%
%%
As a typical massive system, image restoration based on the 
Markov random field (MRF) model has been investigated by
statistical mechanical technique of disordered spin systems 
\cite{PB95,NW99,CI99,Inoue2001}. 
Among these results, statistical mechanical analysis 
succeeded in evaluating measure of successfulness for 
image restoration and made their hyper-parameter dependence 
clear \cite{NW99,CI99,Inoue2001}. 
However, all of those researches were restricted to 
studies of static properties of image  
restoration. In the context of Bayesian statistical approach, 
we usually use the Markov chain Monte Carlo (MCMC) method 
to obtain {\it maximum a posteriori} (MAP) estimate by 
simulated annealing \cite{Geman}, or to calculate expectations  
over posterior distribution for 
{\it maximum posterior marginal} (MPM) estimation \cite{Marroquin}. 
In the recent study by Nishimori and Wong \cite{NW99}, 
they introduced an infinite 
range mean-field version of the MRF model and calculated  
the overlap between the original image and restored one 
analytically. 
However, they did not investigate the dynamical process 
of image restoration, that is to say, 
the process of the MCMC method by Glauber dynamics 
to obtain the MPM estimate. 
Although it is worth while to 
investigate such dynamical processes 
in image restoration, 
relatively little progress has been made in 
the theoretical understanding of them. 
Recently,  Inoue and Carlucci \cite{Inoue2001} 
investigated dynamical properties of 
gray-scale image restoration 
using the mean-field Q-Ising spin glass model analytically. 
They found that 
the MPM estimate gets worse than 
the degraded image when one fails to set the 
hyper-parameters appropriately. 
Therefore, it is important to study  
how we should infer the optimal hyper-parameters. 
As an approach to estimate the optimal hyper-parameters, 
{\it maximum marginal likelihood} (MML) method has been 
used by many authors in practical situations 
\cite{GemanMcClure,Tanaka01}. 
If one maximizes the marginal likelihood by gradient descent, 
Boltzmann machine-type learning equations are obtained 
and these equations contain 
expectations over both posterior and prior 
distributions. 
In order to carry out those expectations, 
we usually use the MCMC method.  
However, it is hard to evaluate the performance 
of the MML estimation due to difficulties to 
simulate the thermo-dynamically equilibrium 
state within reliable precision. 
Therefore, we need some analytical and rigorous 
studies on the hyper-parameter estimation. 
Obviously, the learning process of the hyper-parameter 
estimation and 
the stochastic process of the MCMC method as {\it dynamics}. 
From a view point of 
statistical mechanics of spin systems, 
the process of the hyper-parameter estimation 
is regarded as a dynamics of spin 
system in which coupling constant 
and field strength 
are time dependent variables. 
Then, time dependence of these 
variables is determined by the 
algorithm we choose to maximize 
the marginal likelihood. 
As far as we know, 
no studies have ever tried to 
investigate those dynamical properties 
analytically.   
In this paper, we investigate dynamical properties of 
image restoration including hyper-parameter estimation by using 
statistical mechanical technique.

This paper is organized as follows. 
In the next section, according to Nishimori and Wong \cite{NW99}, 
we explain statistical mechanical formulation of image 
restoration in the context of the MPM estimation.
 
In Sec. \ref{sec3},  we derive differential equations 
with respect to macroscopic observables 
of the infinite range mean-field MRF model 
from microscopic Master equation. 
By solving these differential equations,  
we discuss relaxation process of image restoration. 
In Sec. \ref{sec4}, marginal likelihood as 
a function of hyper-parameters is calculated 
by replica method. 
We also derive Boltzmann machine-type learning equations 
to maximize the marginal likelihood by gradient descent. 
Flows in hyper-parameter space are obtained by 
analyzing the learning equations.  
In the same section, we investigate 
the performance of 
EM (expectation and maximization) algorithm \cite{Dempster} 
which is widely used to estimate hyper-parameters from 
incomplete data sets. 
It is well known that 
EM algorithm shows faster convergence 
at the beginning of the algorithm 
than some other algorithm does.  
However, there is no study to make this property  
clear by using some solvable models.   
In this section, 
we compare the performance of 
EM algorithm with that of gradient descent explicitly. 
The final section is devoted to summary.

%%%%%%%%%%%%%%%%%%%%%%%%%%%%%%%%%%%%%%%%%%%%%%%%%%%%%%%%%%%%%%%%%%%%%%%
\section{Statistical mechanical formulation for image restoration}
\label{sec2}
%%%%%%%%%%%%%%%%%%%%%%%%%%%%%%%%%%%%%%%%%%%%%%%%%%%%%%%%%%%%%%%%%%%%%%%

In this section, we explain how we formulate 
image restoration as a problem of disordered spin system. 
%%%%%%%%%%%%%%%%
According to Nishimori and Wong \cite{NW99}, 
we consider black and while image. 
Then, an original image is denoted by $N$-dimensional 
vector $\{\xi\} \,\equiv\,
(\xi_{1},\xi_{2},{\cdots},\xi_{N})$ 
and each pixel $\xi_{i}$ takes  $\pm 1$. 
These pixels are located on an arbitrary lattice in 
two dimension.  
In order to 
treat image restoration 
by statistical mechanics of disordered spin systems, 
we should assume that the original 
image is given by {\it a priori} Boltzmann-Gibbs distribution 
%%%%%%%%%%%%%%%%%
\begin{eqnarray}
P(\{\xi\}) =     
\frac{{\exp}\left( \beta_{s} \sum_{ij}\xi_{i}\xi_{j}\right)}
{Z_{s}},\,\,\,\,\,
Z_{s} = \sum_{\xi}
{\exp}\left(\beta_{s}
\sum_{ij}\xi_{i}\xi_{j}
\right) 
\label{ferro}
\end{eqnarray}
%%%%%%%%%%%%%%%%%%%%%%%%%%%%%%%%%%%%%
where $\sum_{ij}(\cdots)$ is carried out 
for all nearest neighboring pixels.
Thus, we use 
a snapshot of the MCMC simulation 
for the ferromagnetic Ising model 
as an original image. 
%%%%%%%%
$T_{s}(\equiv \beta_{s}^{-1})$ 
appearing in the argument of the exponential (\ref{ferro}) 
corresponds to temperature.  
We obtain pictures of all black or all white 
when we set $T_s \rightarrow 0$, while  
we obtain random noise pictures 
in the limit of $T_{s} \rightarrow \infty$.  
%%%%%%%%%%%%%%%%%%%%%%%%%%%%%%%
A particular original image  $\{\xi\}$ is 
degraded to a particular damaged picture  
$\{\tau\} \equiv (\tau_{1},\tau_{2},{\cdots},\tau_{N})$ 
by noise channel represented by the following conditional probability  
%%%%%%%%%%%%%%%%%%%%%%%%%%%%%%%
\begin{eqnarray}
P(\{\tau\}|\{\xi\}) & = & 
\frac{{\exp}\left(\beta_{\tau}\sum_{i}\tau_{i}\xi_{i}\right)}
{(2{\cosh}\beta_{\tau})^{N}}
\label{noiseBSC}
\end{eqnarray}
%%%%%%%%%%%%%%%%%%%%%%%%%%%%%%%%%%%%%%%%%%%
where sum $\sum_{i}(\cdots)$ is carried out 
for all pixels and we assumed that 
each pixel is degraded independently. 
$\beta_{\tau}$ represents 
a noise level of the channel 
because the above expression is rewritten as 
$P(-\xi_{i}|\xi_{i})=p=1-P(\xi_{i}|\xi_{i})$ 
with 
$p={\rm e}^{-\beta_{\tau}}/({\rm e}^{\beta_{\tau}}
+{\rm e}^{-\beta_{\tau}})$ for all pixels independently. 
Therefore, 
this kind of noise is referred to as 
{\it Binary Symmetric  Channel} (BSC).

%%%%%%%%%%
The BSC is easily extended 
to the {\it Gaussian Channel} (GC) as follows.   
%%%%%%%%%%%%%%%%%%%%%%%%%%%%%%
\begin{eqnarray}
P(\{\tau\}|\{\xi\}) & = & 
\frac{1}{(\sqrt{2\pi}\tau)^{N}}
{\exp}\left(
-\frac{\sum_{i}(\tau_{i}-\tau_{0}\xi_{i})^{2}}
{2\tau^{2}} 
\right) =  
F_{\rm GC} (\{\tau\})
{\exp}\left(
\frac{\tau_{0}}{\tau^{2}}
\sum_{i}\tau_{i}\xi_{i}
\right) 
\label{GauC1} \\
%%%%%%%%%%%%%%%%%
F_{\rm GC}(\{\tau\})  & \equiv  & 
\frac{1}{(\sqrt{2\pi}\tau)^{N}}
{\exp}\left(
-\frac{\sum_{i}(\tau_{i}^{2}+\tau_{0}^{2})}
{2\tau^{2}}
\right) 
\label{GauC2}
\end{eqnarray}
%%%%%%%%%%%%%%%%%%%%%%%%%%
If we replace $F_{\rm GC} (\{\tau\})$ appearing 
in Eq. (\ref{GauC1}) by 
%%%%%%%%%%%%%%%%%
\begin{eqnarray}
F_{\rm BSC} (\{\tau\}) & \equiv & 
\frac{1}{(2\cosh \beta_{\tau})^{N}}
\prod_{i}\left\{
\delta (\tau_{i}-1) + 
\delta (\tau_{i}+1)
\right\}
\end{eqnarray}
%%%%%%%%%%%%%%
with $\tau_{0}/\tau^{2} = \beta_{\tau}$, 
the BSC (Eq. (\ref{noiseBSC})) is recovered. 
%%%%%%%%%%%%%%%%%%%%%%%%%%%%%%
We should notice that 
a sum $\sum_{\tau} \Omega(\{\tau\})$ 
for an arbitrary 
function $\Omega(\{\tau\})$ is 
calculated in terms of $F_{\rm GC, BSC}(\{\tau\})$ as 
%%%%
\begin{eqnarray}
\sum_{\tau}\Omega(\{\tau\}) & = & 
\int \cdots \int d\{\tau\} F_{\rm GC, BSC}(\{\tau\})\Omega(\{\tau\})
\end{eqnarray}
%%%
where we defined 
$d\{\tau\} \equiv d\tau_{1}d\tau_{2}\cdots d\tau_{N}$. 
%%%%%%%%%%%%%  
Then, Bayes theorem gives the posterior distribution 
%%%%%%%%%%%%%%%%%%%%%%%%%
\begin{eqnarray} 
P(\{\sigma\}|\{\tau\}) & = & 
\frac{P(\{\tau\}|\{\sigma\})P(\{\sigma\})}
{\sum_{\sigma}
P(\{\tau\}|\{\sigma\})P(\{\sigma\})} =  
\frac{{\rm e}^{J\sum_{ij}\sigma_{i}\sigma_{j}
+h\sum_{i}\tau_{i}\sigma_{i}}}
{\sum_{\sigma}
{\rm e}^{J\sum_{ij}\sigma_{i}\sigma_{j}
+h\sum_{i}\tau_{i}\sigma_{i}}}
\label{post00}
\end{eqnarray}
%%%%%%%%%%%%%
where $J$ and $h$ are hyper-parameters and 
we introduced models of the prior (Eq. (\ref{ferro})) 
and the likelihood (Eq. (\ref{noiseBSC})) as  
%%%%%%%%%%%%%%%%%%%%%%
\begin{eqnarray}
P(\{\sigma\}) = \frac{{\exp}
\left(J\sum_{ij}\sigma_{i}\sigma_{j}\right)}
{Z_{\Pi}}, \,\,\,\,\, P(\{\tau\}|\{\sigma\}) =  
\frac{{\exp}\left(
h\sum_{i}\tau_{i}\sigma_{i}
\right)}
{Z_{L}}, 
\label{priandlike}
\end{eqnarray}
%%%%
respectively. 
%%%%%%%%%%%%%%%%%%%%%%%
A configuration 
$\{\sigma\} \equiv 
(\sigma_{1},\sigma_{2}, \cdots, \sigma_{N})$ 
denotes an estimate of a particular 
original image $\{\xi\}$. 
%%%%%%%%%%%%%%%%%%%%%%%%%%%%%%
$Z_{\Pi}$ and $Z_{L}$ in 
Eq. (\ref{priandlike}) 
are normalization constants given by 
%%%%%%%%%%%%%%
\begin{eqnarray}
Z_{\Pi} = \sum_{\sigma}{\exp}\left(
J\sum_{ij}\sigma_{i}\sigma_{j}
\right),\,\,\,\,\, 
Z_{L} = \sum_{\tau}
{\exp}\left(
h\sum_{i}\tau_{i}\sigma_{i}
\right).
\label{normalconsts}
\end{eqnarray}
%%%%%%%%%%%%%%%%%%%
%%%%
It is 
important for us 
to bear in mind that 
$Z_{L}$ is independent of 
$\{\sigma\}$ for 
both the BSC and the GC. 
Actually, $Z_{L}$ leads to 
%%%%%%
\begin{eqnarray}
Z_{L} =  
\int \cdots \int d\{\tau\} F_{\rm BSC}(\{\tau\})\exp\left(
h\sum_{i}\tau_{i}\sigma_{i}\right) = \left(
\frac{2\cosh h}{2\cosh \beta_{\tau}}
\right)^{N}
\end{eqnarray}
%%% 
for the BSC and 
%%%%%
\begin{eqnarray}
Z_{L} = 
\int \cdots \int d\{\tau\} F_{\rm GC}(\{\tau\})\exp\left(
h\sum_{i}\tau_{i}\sigma_{i}\right) = 
\exp \left(
-\frac{N\tau_{0}^{2}}{2\tau^{2}}
+\frac{N\tau^{2}h^{2}}{2}
\right)
\label{GauC3}
\end{eqnarray}
%%%
for the GC. 
%%%

In the context of MAP estimation, we
choose the estimate $\{\sigma\}$ as a 
grand state of the following Hamiltonian 
(cost function) 
%%%%%%%%%%%%%%%%%%%%%
\begin{eqnarray}
{\cal H}(\{\sigma\}) & = & 
-J\sum_{ij}\sigma_{i}\sigma_{j}
-h\sum_{i}\tau_{i}\sigma_{i}.
\end{eqnarray}
%%%%%%%%%%%%%%%%%%%%%%%%%
In order to obtain the 
grand state, we usually use 
simulated annealing \cite{Kirkpatrick83} or 
mean field annealing \cite{Geiger91}. 
%%%%%%%%%%%%%%%%%%

On the other hand, 
in the context of MPM estimation, we
first calculate the marginal distribution 
around a single pixel $\sigma_{i}$ :  
%%%%%%%%%%%%%
\begin{eqnarray}
P(\sigma_{i}|\{\tau\}) & = & 
\sum_{\{ \sigma \} \neq \sigma_i}
P(\{\sigma\}|\{\tau\})
\end{eqnarray}
%%%%%%%%%%%%%%%%%%%%%%%%%%%%%%%%%%%
and we choose the 
sign of the difference 
between $P(\sigma_{i}=+1|\{\tau\})$ and 
$P(\sigma=-1|\{\tau\})$ as an 
estimate of the $i$-th pixel $\hat{\xi}_{i}$ as 
%%%%%%%%%%%%%%%%
\begin{eqnarray}
\hat{\xi}_{i}  & = &  
{\arg} \displaystyle{\max_{\sigma_{i}}}P(\sigma_{i}|\{\tau\})= 
{\rm sgn}\left(
\sum_{\sigma_{i}=\pm 1}
P(\sigma_{i}|\{\tau\})
\right) = {\rm sgn}
\left(
\frac{\sum_{\sigma}\sigma_{i}
P(\{\sigma\}|\{\tau\})}
{\sum_{\sigma}
P(\{\sigma\}|\{\tau\})} 
\right) \nonumber \\
\mbox{} & \equiv &    
{\rm sgn}\left(
\langle \sigma_{i} \rangle_{J,h}
\right).
\end{eqnarray}
%%%%%%%%%%%%%%%%%%%%%%%%%%%
In this expression, we defined 
$\langle \sigma_{i} \rangle_{J,h}$ as 
an average of the $i$-th pixel $\sigma_{i}$ 
over the posterior distribution 
(\ref{post00}) and this is written 
explicitly as 
%%%%%%%%%%%%%
\begin{eqnarray}
\langle \sigma_{i} \rangle_{J,h} & = & 
\frac{\sum_{\sigma}
\sigma_{i}{\rm e}^{J\sum_{ij}\sigma_{i}\sigma_{j}
+h\sum_{i}\tau_{i}\sigma_{i}}
}
{\sum_{\sigma}
{\rm e}^{J\sum_{ij}\sigma_{i}\sigma_{j}
+h\sum_{i}\tau_{i}\sigma_{i}}
}.
\end{eqnarray}
%%%%%%%%%%%%%%%%%
This corresponds to a local magnetization 
of the spin system that is described by 
the Hamiltonian ${\cal H}(\{\sigma\})$ 
at temperature $T=1$. 
Thus, in order to investigate properties of 
the MPM estimation for image restoration,  we should 
study the random field Ising model 
described by ${\cal H}(\{\sigma\})$. 
Then, we are interested in the quantity : 
%%%%%%%%%%%%%%
\begin{eqnarray}
M(J,h) & \equiv  & \sum_{\xi, \tau}
P(\{\xi\})P(\{\tau\}|\{\xi\})
\,\xi_{i}\,{\rm sgn}(\langle \sigma_{i} \rangle_{J,h})
\end{eqnarray}
%%%%%%%%%%%%%%
which means the averaged overlap between an arbitrary 
original pixel $\xi_{i}$ and the MPM estimate 
$\hat{\xi}_{i}={\rm sgn}(\langle \sigma_{i} \rangle_{J,h})$. 
Apparently, the best restoration of the 
original image  
is achieved when the overlap $M$ is as close to $1$ as 
possible. For this averaged 
overlap $M(J,h)$, the next inequality holds \cite{NW99}. 
%%%%%%%%%%%%%
\begin{eqnarray}
M(J,h) & \leq & M(\beta_{s},\beta_{\tau}) 
\end{eqnarray}
%%%%%%%%%%%%%%%%%%
This inequality means that the averaged 
overlap $M$ takes its maximum 
when one sets the hyper-parameters to 
their true values, namely, $J=\beta_{s}$ and 
$h=\beta_{\tau}$. 
However, 
it is impossible to derive the hyper-parameter dependence 
of the overlap around its optimal value 
$M(\beta_{s},\beta_{\tau})$ from the above inequality. 
To investigate this dependence, Nishimori and 
Wong \cite{NW99} introduced a mean-field 
version of the MRF model and calculated  
the overlap as a function of $J$ and $h$.  
The mean-field model is rather an artificial 
model in which every pixel is connected to the others, 
however, this model is very useful to 
discuss the behavior of macroscopic quantities 
of the system, like the overlap $M$. 
Using replica method \cite{SK75}, 
one obtains saddle point equations 
%%%%%%%%%%%%%%%%%%%%%%%%
\begin{eqnarray}
m_{0} & \equiv  & \frac{1}{N}\sum_{i}\xi_{i}={\tanh}(\beta_{s} m_{0}) 
\label{m0} \\
%%%%%%%%%%%%
m & \equiv & \frac{1}{N}\sum_{i}\sigma_{i}=
\frac{\sum_{\xi}{\rm e}^{\beta_{s}m_{0}\xi}}
{2{\cosh}(\beta_{s}m_{0})}
\int_{-\infty}^{\infty}
Dx\, {\tanh}(Jm+\tau h x+\tau_{0}h\xi) 
\label{m} \\
%%%%%%%%%%%%%%%
M & \equiv & 
\frac{1}{N}\sum_{i}\xi_{i}\hat{\xi}_{i}= 
\frac{\sum_{\xi}{\rm e}^{\beta_{s}m_{0}\xi}}
{2{\cosh}(\beta_{s}m_{0})}
\int_{-\infty}^{\infty}
Dx\, \xi {\rm sgn}(Jm+\tau h x +\tau_{0}h\xi)
\label{M}
\end{eqnarray}
%%%%%%%%%%%%%%%%%%%%%
where we defined Gaussian integral measure 
by $Dx \equiv dx \, {\rm e}^{-x^2/2}/\sqrt{2\pi}$. 
Equation (\ref{m0}) determines macroscopic 
properties of the original image 
given by the 
Hamiltonian $-\sum_{ij}\xi_i \xi_j$ 
at temperature 
$T_{s}(\equiv \beta_{s}^{-1})$. 
%%%%%%%%%%%%
From statistical mechanical point of view, 
$m_{0}$ corresponds to magnetization 
of the mean-field ferro-magnetic Ising model. 
For a given $T_{s}$, 
one obtains $m_{0}$ by solving Eq. (\ref{m0}).  
Substituting $T_{s}$, $m_{0}$ 
and noise parameters $\tau_{0}$ (a center of Gaussian) and 
$\tau$ (a standard deviation) into Eq.  (\ref{m}), one obtains 
magnetization $m$ for the restored image 
system $\{\sigma\}$ as a function of $T_{m}(\equiv J^{-1})$ 
and $h$. 
Then, one substitutes $m(T_{m},h)$ into the 
expression of $M$, and finds hyper-parameter dependence 
of the overlap explicitly. 
In FIG. \ref{fig1}, we 
plot the overlap $M$ as a function of $1/J (\equiv T_m)$.  
We set $\tau=\tau_{0}=1$ ($\beta_{\tau}=\tau_{0}/\tau^{2}=1$) and 
temperature of the original image is chosen as $T_{s}=0.9$. 
The overlap for the two cases of the field $h$, namely, 
$h=\beta_{\tau}T_{s}J=\tau_{0}T_{s}J/\tau^{2}=0.9J\equiv 
h_{\rm opt}$ {\sf (a)} and $h=1$ {\sf (b)} are shown.  
We should notice that 
the MAP estimate is obtained in the 
limit of $T_m \rightarrow 0$ 
keeping the ratio $h/J$ constant.  
Therefore, the overlap for the MAP 
estimate depends on the ratio $h/J$ and 
takes its maximum when we set $h/J=\beta_{\tau}T_s=0.9$ 
(see FIG. \ref{fig1}  {\sf (a)}).
From this figure, we see that the overlap takes its 
maximum at $T_{m}=T_{s}=0.9$ and $h=\beta_{\tau}=\tau_{0}/\tau^{2}=1$. 
In the next section, we focus our attention on 
dynamics of the MPM estimation.

%%%%%%%%%%%%%%%%%%%%%%%%%%%%%%%%%%%%%%%%%%%%%%%%%%%%%%%%%%%%%%%%%%%%%%%
\section{Dynamics of image restoration}
\label{sec3}
%%%%%%%%%%%%%%%%%%%%%%%%%%%%%%%%%%%%%%%%%%%%%%%%%%%%%%%%%%%%%%%%%%%%%%%

In the previous section, we showed the 
performance of the MPM estimation. 
However, in those calculations, we assumed that the 
system already reached the equilibrium state. 
In other words, each state $\{\sigma\}$ 
obeys the Boltzmann-Gibbs distribution 
${\sim}\,{\rm e}^{-{\cal H}(\{\sigma\})}$. 
When we need to generate the 
distribution 
to calculate the MPM estimate 
${\rm sgn}(\langle \sigma_{i} \rangle_{J,h})$, 
we often use the MCMC method and simulate 
the equilibrium states on computer. 
Therefore, 
it is important to study how the system 
relaxes to its equilibrium state and 
grasp the behavior of time evolutionary observables analytically. 
As far as we know, there is no research to deal with 
dynamics of image restoration including 
hyper-parameter estimation analytically. 
%%%%%%%%%%%%
In this section, for 
the infinite range mean-field MRF model, we derive 
differential equations 
with respect to macroscopic order parameters of 
the restored image system from microscopic 
Master equation.

First of all, we should 
remember that a transition rate  
$w_{k}(\{\sigma\})$ from 
$\{\sigma\}\,\equiv\,(\sigma_{1},\sigma_{2},
{\cdots},\sigma_{k},\cdots,\sigma_{N})$ to 
$\{\sigma\}^{'} \,\equiv\,
(\sigma_{1},\sigma_{2},\cdots,-\sigma_{k},{\cdots},\sigma_{N})$ 
leads to 
%%%%%%%%%%%%%%%%%%%
\begin{eqnarray}
w_{k}(\{\sigma\}) =  
\frac{1}{2}\left[
1-\sigma_{k}\tanh [h_{k}(\{\sigma\})]
\right],\,\,\,\,\,\,
h_{k}(\{\sigma\}) =  
\frac{J}{N}\sum_{j}\sigma_{j}+h\tau_{k}
\end{eqnarray}
%%%%%%%%%%
in the context of the Glauber dynamics of the MCMC method. 
%%%%
It is important 
for us to bear in mind that Hamiltonian ${\cal H}(\{\sigma\})$ 
of the system is rewritten in terms of 
$h_{k}(\{\sigma\})$ as 
%%%%%%%%%%%%%%
\begin{eqnarray}
{\cal H}(\{\sigma\}) & = & -\sum_{k}h_{k}(\{\sigma\})\sigma_{k}
\end{eqnarray}
%%%%%%%%%%%%%%%%%%
where we rescaled the coupling $J$ as $J/N$ to 
take a proper thermo-dynamic limit (Hamiltonian 
should be of order $N$).  

Then, probability $p_{t}(\{\sigma\})$ 
that the system visits a state $\{\sigma\}$ 
at time $t$ obeys the Master equation 
%%%%%%%%%%%%%%%%%%%
\begin{eqnarray}
\frac{dp_{t}(\{\sigma\})}{dt} & = & 
\sum_{k=1}^{N}\left[
p_{t}(F_{k}(\{\sigma\}))w_{k}(F_{k}(\{\sigma\}))-
p_{t}(\{\sigma\})w_{k}(\{\sigma\})
\right]
\label{dpdt}
\end{eqnarray}
%%%%%%%%%%%
where we defined single spin flip operator $F_{k}$ by 
%%%%%%%%%%%%%
\begin{eqnarray}
F_{k}(\{\sigma\}) & = & (\sigma_{1},\sigma_{2},
{\cdots},-\sigma_{k},{\cdots},\sigma_{N}) =\{\sigma\}^{'}.
\end{eqnarray}
%%%%%%%%%%%%%%%%%%
Distribution $P_{t}(m,a)$,  which is 
probability that the system has macroscopic order parameters
%%%%%%%%%%%%%%%%%%
\begin{eqnarray}
m (\{\sigma\}) \equiv   \frac{1}{N}\sum_{i}\sigma_{i},\,\,\,\,\,\,
a (\{\sigma\}) \equiv   \frac{1}{N}\sum_{i}\tau_{i}\sigma_{i}
\end{eqnarray}
%%%%%%%%%%%%
at time $t$,  is written in terms of the distribution 
$p_{t}(\{\sigma\})$ of the microscopic state $\{\sigma\}$ as 
%%%%%%%%%%%%%%
\begin{eqnarray}
P_{t}(m, a) & = & \sum_{\sigma}
p_{t}(\{\sigma\})\delta(m-m(\{\sigma\})) 
\delta(a-a(\{\sigma\}))
\end{eqnarray}
%%%%%%%%%%%
where $\delta(\cdots)$ is a delta function.
%%%%%%%%
Taking a derivative of $P_{t}(m,a)$ with respect to $t$ and 
substituting Eq. (\ref{dpdt}) into this expression 
and making a Taylor expansion in powers of 
$2\sigma_{k}/N$ and $2\tau_{k}\sigma_{k}/N$ 
(so-called {\it Kramers-Moyal expansion}), 
we obtain 
%%%%%%%%%%
\begin{eqnarray}
\mbox{} & \mbox{} & \frac{dP_{t}(m,a)}{dt}  =  
\frac{\partial}{\partial m}
P_{t}(m,a)\left\{
m-\frac{\sum_{\xi}{\rm e}^{\beta_{s}m_{0}\xi}}
{2{\cosh}(\beta_{s}m_{0})}
\int_{-\infty}^{\infty}
Dx \,{\tanh}(Jm+h\tau x +h\tau_{0}\xi)
\right\} \nonumber \\
\mbox{} & + & 
\frac{\partial}{\partial a}
P_{t}(m,a)
\left\{
a-\frac{\sum_{\xi}{\rm e}^{\beta_{s}m_{0}\xi}}
{2{\cosh}(\beta_{s}m_{0})}
\int_{-\infty}^{\infty}
Dx (\tau x+\tau_{0}\xi){\tanh}
(Jm+h\tau x +h\tau_{0}\xi)
\right\} \nonumber \\
\mbox{} & + & {\cal O}(N^{-1}).
\label{dpdt2}
\end{eqnarray} 
%%%%%%%%%%%%%%
Thus, we derived the time dependent distribution 
of macroscopic quantities from the microscopic 
Master equation Eq. (\ref{dpdt}). 
Finally, we construct differential 
equations with respect to the macroscopic quantities $m$ and $a$. 
%%%%%%%%%%%%
Substituting a form of distribution 
%%%%%%%%
\begin{eqnarray}
P_{t}(m,a) & = & {\delta}(m-m(t)){\delta}(a-a(t))
\end{eqnarray}
%%%%%%%%%%%%
into Eq. (\ref{dpdt2}) and  
calculating some integrals, we obtain 
%%%%%%%%%%%%%%%
\begin{eqnarray}
\frac{dm}{dt} & = & 
-m +\frac{\sum_{\xi}
{\rm e}^{\beta_{s}m_{0}\xi}}{2{\cosh}(\beta_{s}m_{0})}
\int_{-\infty}^{\infty}
Dx \,{\tanh}(Jm+h\tau x+h\tau_{0}\xi)
\label{dmdt} \\
%%%%%%%%%%%%%%%%%%%
\frac{da}{dt} & = & 
-a +  \frac{\sum_{\xi}
{\rm e}^{\beta_{s}m_{0}\xi}}{2{\cosh}(\beta_{s}m_{0})}
\int_{-\infty}^{\infty}
Dx (\tau x +\tau_{0}\xi){\tanh}(Jm+h\tau x+h\tau_{0}\xi) 
\label{dadt}.
\end{eqnarray}
%%%%%%%%%%
These two equations 
describe a relaxation of the system to the 
equilibrium state. 
%%%%%%%%
We should notice that 
order parameter $a$ is a slave variable 
in the sense that the order parameter $m$ relaxes independently, 
whereas the relaxation of $a$ depends on $m$. 
Therefore, the behavior of $a$ is completely determined by $m$. 
%%%%%%%%%%%%
For this reason, from now on, we 
disregard  Eq. (\ref{dadt}). 
%%%%%%%%%%

It is easy to see that 
in the limit of $t\rightarrow \infty$ and 
$dm/dt=0$, 
the saddle point equation (\ref{m}) is recovered. 
%%%%%%%%%%%
As the overlap $M$ is written 
in terms of $m$ [see Eq. (\ref{M})], 
time evolution of the overlap is 
obtained by substituting the time dependence 
of the magnetization $m(t)$ into the 
expression of  $M$. 
%%%%%%%%%%%%%%%%%%%%%%%%%%%%%%%%%%%%%

Using the same technique as the 
procedure to derive the 
differential equation with 
respect to $m$, 
the differential equation for 
the magnetization $m_{1}$ of the prior system 
$P(\{\sigma \})={\exp}(J\sum_{ij}\sigma_{i}\sigma_{j})/
\sum_{\sigma}\exp(J\sum_{ij}\sigma_{i}\sigma_{j})$ 
is obtained as 
%%%%%%%%%%%%%%%%%
\begin{eqnarray}
\frac{dm_{1}}{dt} & = & 
-m_{1}+\tanh (m_{1}J)
\label{dm1dt}.
\end{eqnarray}
%%%%%%%%%%%%%%%%%
%%%%%%%%%%%%%%
Although in these equations 
we regard the hyper-parameters $J$ and $h$ as 
constant variables, one should treat them as 
time dependent parameters, 
that is, $J(t)$ and $h(t)$ 
from a view point of hyper-parameter estimation. 
Of course, details of the time dependence of 
$J(t)$ and $h(t)$ 
depend on a particular algorithm 
of hyper-parameter estimation. 
In the next section, 
we investigate properties of hyper-parameter 
estimation as dynamical process of 
coupling constant $J(t)$ and field strength $h(t)$.

%%%%%%%%%%%%%%%%%%%%%%%%%%%%%%%%%%%%%%%%%%%%%%%%%%%%%%%%%%%%%%%%%%%%%%%
\section{Hyper-parameter estimation}
\label{sec4}
%%%%%%%%%%%%%%%%%%%%%%%%%%%%%%%%%%%%%%%%%%%%%%%%%%%%%%%%%%%%%%%%%%%%%%%

In the previous two sections, we investigated 
both static and dynamical properties of image restoration. 
From those results, we obtained hyper-parameter 
dependence of the overlap explicitly. 
Moreover, 
for a particular constant hyper-parameter set $(J,h)$,  
we derived the differential equations 
which describe the relaxation of the system. 
As one of the authors reported in 
\cite{Inoue2001}, 
if one fails to set the hyper-parameters appropriately, 
the restored image gets worse than the degraded image. 
In practical situations, we do not know 
the optimal value of the hyper-parameters  
before we carry out the MCMC simulations. 
Therefore, we need to 
determine the optimal value by using 
only information about 
the degraded image $\{\tau\}$. 
Of course, 
it is possible for us to 
construct some robust algorithms for 
hyper-parameter tuning 
and several authors 
reported such algorithms 
based on {\it selective freezing} \cite{Wong2000} or 
{\it quantum fluctuation} \cite{Inoue2001b}. 
However, if one seeks 
the optimal restoration, 
hyper-parameter estimation becomes a very important problem.

About ten years ago, Iba \cite{Iba} studied  
the performance of the MML method with the assistance of the 
MCMC simulations for 
the same problem as ours. 
%%%%
However, as he mentioned in his paper, 
the results are not enough to make its performance 
clear due to the difficulties of 
simulating the equilibrium 
state within reliable precision. 
With this fact in mind, in this section, 
we calculate marginal likelihood as a 
function of hyper-parameters 
analytically. 
From the marginal 
likelihood, we derive Boltzmann machine-type 
learning equations and investigate their behavior 
quantitatively.

%%%%%%%%%%%%%%%%%%%%%%%%%%%%%%%%%%%%%%%%%%%%%%%%%%%%%%%%%%%%%%%%%%%
\subsection{Maximum marginal likelihood method}
%%%%%%%%%%%%%%%%%%%%%%%%%%%%%%%%%%%%%%%%%%%%%%%%%%%%%%%%%%%%%%%%%%%%%

In statistics, maximum marginal likelihood 
(MML) method is used to infer hyper-parameters  
appearing in the posterior distribution \cite{PB95,GemanMcClure,Zhou97}. 
In the context of image restoration, 
marginal likelihood (logarithm of marginal likelihood) 
is given by 
%%%%%%%%%
\begin{eqnarray}
-K(J,h : \{\xi, \tau\}) & \equiv &  
{\log}\sum_{\sigma}P(\{\tau\}|\{\sigma\})P(\{\sigma\}) \nonumber \\
\mbox{} & = & 
{\log}\left(
\sum_{\sigma}{\rm e}^{J\sum_{ij}\sigma_{i}\sigma_{j}
+h\sum_{i}\tau_{i}\sigma_{i}}
\right)-\log Z_{\Pi}-\log Z_{L} 
\label{margl}
\end{eqnarray}
%%%%%%%%%%%
where $Z_{\Pi}$ and $Z_{L}$ are 
given by Eq. (\ref{normalconsts}). 
We should remember that 
$Z_{L}$ is independent of 
$\{\sigma\}$ for both cases of the 
BSC and the GC. 
%%%%%%%
Usually, 
we attempt to maximize the 
marginal likelihood by using 
gradient descent with respect to  
$J$ and $h$. 
This result leads to 
the following Boltzmann machine-type 
learning equations 
%%%%%%%%%%%%%%%%%%%%
\begin{eqnarray}
c_{J}\frac{dJ}{dt} & = &  
-\frac{\partial K(J,h : \{\xi,\tau\})}{\partial J}  =
\frac{\sum_{\sigma}(\sum_{ij}\sigma_{i}\sigma_{j})
{\rm e}^{J\sum_{ij}\sigma_{i}\sigma_{j}
+h\sum_{i}\tau_{i}\sigma_{i}}}
{\sum_{\sigma}{\rm e}^{J\sum_{ij}\sigma_{i}\sigma_{j}
+h\sum_{i}\tau_{i}\sigma_{i}}} -  
\frac{\sum_{\sigma}(\sum_{ij}\sigma_{i}\sigma_{j})
{\rm e}^{J\sum_{ij}\sigma_{i}\sigma_{j}}}
{\sum_{\sigma}{\rm e}^{J\sum_{ij}\sigma_{i}\sigma_{j}}} 
\label{dJdt} \\
%%%%%%%%%%%%%%%%%%%%%
c_{h}\frac{dh}{dt} & = &  
-\frac{\partial K(J,h : \{\xi,\tau\})}{\partial h} = 
\frac{\sum_{\sigma}(\sum_{i}\tau_{i}\sigma_{i})
{\rm e}^{J\sum_{ij}\sigma_{i}\sigma_{j}
+h\sum_{i}\tau_{i}\sigma_{i}}}
{\sum_{\sigma}{\rm e}^{J\sum_{ij}\sigma_{i}\sigma_{j}
+h\sum_{i}\tau_{i}\sigma_{i}}}-
\frac{\partial {\log}Z_{L}}{\partial h}
\label{dhdt}
\end{eqnarray}
%%%%%%%%%%%%%
where $c_{J}$ and $c_{h}$ are relaxation times. 
%%%%%%
Thus, by solving these two equations, 
we maximize the marginal likelihood $-K(J,h : \{\xi,\tau\})$ and 
obtain the values of hyper-parameters 
as a fixed point of the equations. 
%%%%%%%%%%%%%%%
Then, we should notice that 
these two equations contain expectations of the 
quantities $\sum_{ij}\sigma_{i}\sigma_{j}$ and 
$\sum_{i}\tau_{i}\sigma_{i}$ over the posterior 
and the prior distributions. 
Therefore, when we solve Eqs. (\ref{dJdt}), (\ref{dhdt}) 
numerically, we should calculate these expectations 
at each time step of the Euler method. 
Iba \cite{Iba} carried out 
the MCMC method to calculate the expectations and 
evaluated time dependence of the 
hyper-parameters $J$ and $h$ numerically. 
However, 
the accuracy of his computer simulation is 
not reliable because the time to 
simulate the equilibrium state is not enough.   
%%%%%
Accordingly, it is worth while to 
investigate the performance of the MML method 
analytically using the solvable model. 
In this subsection, 
we use the infinite range mean-field MRF model and 
solve the learning equations (\ref{dJdt}),(\ref{dhdt}) 
exactly.
%%%%%%%%%%%

As our interest is averaged performance of the 
MML method, we should calculate 
the averaged marginal likelihood : 
%%%%%%%%%%%%%%%
\begin{eqnarray}
-\left[
K(J,h : \{\xi,\tau\})
\right]_{\{\xi,\tau\}} & = &   
\left[
{\log}\sum_{\sigma}{\rm e}^{\frac{J}{N}\sum_{ij}
\sigma_{i}\sigma_{j}
+h\sum_{i}\tau_{i}\sigma_{i}}
\right]_{\{\xi,\tau\}} 
-
\left[
{\log}\sum_{\sigma}
{\rm e}^{\frac{J}{N}\sum_{ij}\sigma_{i}
\sigma_{j}}
\right]_{\{\xi,\tau\}} \nonumber \\
\mbox{} & - &   
\left[
{\log}Z_{L}
\right]_{\{\xi,\tau\}}
\label{marginal-like}
\end{eqnarray}
%%%%%%%%%%%%%%%%
where the 
bracket $[\cdots]_{\{\xi,\tau\}}$ means 
the average over the distribution 
$P(\{\tau\}|\{\xi\})P(\{\xi\})$ 
and the sum $\sum_{ij}(\cdots)$ should be 
carried out for all pairs of pixels. 
%%%%%%%%%%%
We should keep in mind that 
we rescaled the coupling constant 
as $J/N$ to make the averaged marginal likelihood 
(difference of free energy) of order $N$.   
In general, it is hard to carry out this 
kind of average, namely, 
$[{\log}Z]_{\{\xi,\tau\}}$. 
%%%%%%%%%%%%%%%%%%%
Then, we replace the average 
with an average of the $n$-th moment of 
$Z$, that is, $Z^{n}$ by using 
%%%%%%%%%
\begin{eqnarray}
[{\log}Z]_{\{\xi,\tau\}} & = & 
\displaystyle{\lim_{n\rightarrow 0}}
\frac{[Z^{n}]_{\{\xi,\tau\}}-1}{n}.
\end{eqnarray}
%%%%%%%
This is refereed to as {\it replica method} \cite{SK75}. 
By using the replica method, we obtain 
the averaged marginal likelihood per pixel as 
%%%%%%%%%%%%
\begin{eqnarray}
-\frac{[K(J,h : \{\xi,\tau\})]_{\{\xi,\tau\}}}
{N} & = & 
-\frac{J}{2}m^{2}+\frac{\sum_{\xi}
{\rm e}^{\beta_{s}m_{0}\xi}}
{2{\cosh}(\beta_{s}m_{0})}
\int_{-\infty}^{\infty}
Dx \,{\log}\,2{\cosh}
(Jm+h\tau x+h\tau_{0} \xi)  \nonumber \\
\mbox{} & + & \frac{J}{2}m_{1}^{2} - {\log}\,
2{\cosh}(m_{1}J)+\frac{\tau_{0}}{2\tau^{2}}
-\frac{\tau^{2}h^{2}}{2}\,\equiv\, -K(J,h)
\end{eqnarray}
%%%%%%%%%%%
where $m$ and $m_{1}$ are 
magnetizations of the spin systems described 
by the posterior and the prior, respectively. 
It should be noticed that as we used the 
GC (Eqs. (\ref{GauC1}), (\ref{GauC2})), the average 
$[\log Z_{L}]_{\{\xi,\tau\}}$ simply 
led to $(Nh^{2}/2)-(N\tau_{0}/2\tau^{2})$ 
(see Eq. (\ref{GauC3})). 

In FIG. \ref{fig2}, 
we plot the averaged marginal likelihood  
as a function of $J$ and $h$. 
%%%%%%%%%%%%%%%%%%%%%
In this figure, 
we see that the averaged marginal likelihood 
takes its maximum when we choose the 
hyper-parameters $(J,h)$ so as to be 
identical to their true values 
$(\beta_{s}=1/T_s = 1.1,  \beta_{\tau}=\tau_{0}/\tau^{2}=1)$ 
(we set $\tau_{0}=\tau=1, T_s =0.9$). 
%%%%%%%%%%%%
This fact is easily checked by the following 
inequality \cite{Iba1999} 
%%%%%%%%%%%%%%%%%%%%%
\begin{eqnarray}
\mbox{} & \mbox{} & 
\{-[K(\beta_{s},\beta_{\tau} : \{\xi,\tau\})]_{\{\xi,\tau\}} \} -
\{-[K(J,h : \{\xi,\tau\})]_{\{\xi,\tau\}} \}  \nonumber \\
\mbox{} & = & 
\sum_{\xi,\tau}
P_{\beta_{\tau}}(\{\tau\}|\{\xi\})
P_{\beta_{s}}(\{\xi\})
{\log}\sum_{\sigma}P_{\beta_{\tau}}(\{\tau\}|\{\sigma\})
P_{\beta_{s}}(\{\sigma\})  \nonumber \\
\mbox{} & - & 
\sum_{\xi,\tau}
P_{\beta_{\tau}}(\{\tau\}|\{\xi\})
P_{\beta_{s}}(\{\xi\})
{\log}\sum_{\sigma}P_{h}(\{\tau\}|\{\sigma\})
P_{J}(\{\sigma\}) \nonumber \\
\mbox{} & = &  
\sum_{\tau}
P_{\beta_{s},\beta_{\tau}}(\{\tau\})
{\log}(P_{\beta_{s},\beta_{\tau}}(\{\tau\})/
P_{J,h}(\{\tau\})) \geq 0
\end{eqnarray}
%%%%%%%%%
where we used non-negativity of {\it Kullback-Libeler information} 
and we defined 
%%%%%%%%%%%%%%%
\begin{eqnarray}
P_{X}(\{\tau\}|\{\sigma\}) \equiv  
\frac{{\exp}(X\sum_{i}\tau_{i}\sigma_{i})}
{\sum_{\tau}{\exp}(X\sum_{i}\tau_{i}\sigma_{i})},\,\,\,\,
P_{X}(\{\tau\}|\{\xi\})  \equiv 
\frac{{\exp}(X\sum_{i}\tau_{i}\xi_{i})}
{\sum_{\tau}{\exp}(X\sum_{i}\tau_{i}\xi_{i})}
\end{eqnarray}
%%%%%%%%%%%%%%%%
\begin{eqnarray}
P_{Y}(\{\sigma\}) \equiv  
\frac{{\exp}(Y\sum_{ij}\sigma_{i}\sigma_{j})}
{\sum_{\sigma}{\exp}(Y\sum_{ij}\sigma_{i}\sigma_{j})},\,\,\,\,
P_{Y}(\{\xi\}) \equiv  
\frac{{\exp}(Y\sum_{ij}\xi_{i}\xi_{j})}
{\sum_{\xi}{\exp}(Y\sum_{ij}\xi_{i}\xi_{j})}
\end{eqnarray}
%%%%%%%%%%%%%
\begin{eqnarray}
P_{X,Y}(\{\tau\}) \equiv 
\sum_{\sigma}P_{X}(\{\tau\}|\{\sigma\})P_{Y}(\{\sigma \})=
\sum_{\xi}P_{X}(\{\tau\}|\{\xi\})P_{Y}(\{\xi\}).
\end{eqnarray}
%%%%%%%%%%%%%%%
Thus, 
we confirm that our mean-field model is not against this 
general inequality. 
We should mentioned that 
static properties of the hyper-parameter 
estimation was investigated by several authors 
using the generalized Gaussian model \cite{TI01}, 
mean-field approximation \cite{PB95} and 
cluster variation method \cite{Tanaka01}. 

For the marginal likelihood (\ref{marginal-like}), 
averaged learning equations with respect to $J$ and $h$ 
are obtained by gradient descent : 
%%%%%%%%%
\begin{eqnarray}
c_{J}\frac{dJ}{dt} =
-\left[
\frac{\partial K(J,h : \{\xi,\tau\})}{\partial J}
\right]_{\{\xi,\tau\}},\,\,\,\,\,
c_{h}\frac{\partial h}{\partial t} =  
-\left[
\frac{\partial K(J,h : \{\xi,\tau\})}{\partial h}
\right]_{\{\xi,\tau\}}.
\end{eqnarray}
%%%%%
The right hand sides of the above 
equations are also evaluated by the replica method.  
%%%%%%
After some algebra, we obtain  
%%%%%%%%%%%%%%%%%
%%%%%%%%%%%%%%%%%%%%%%%%%%
\begin{eqnarray}
c_{J}\frac{dJ}{dt} & = & 
-\frac{m^{2}}{2}+m
\frac{\sum_{\xi}{\rm e}^{\beta_{s}m_{0}\xi}}
{2{\cosh}(\beta_{s}m_{0})}
\int_{-\infty}^{\infty}
Dx\, {\tanh}(Jm+h\tau x+h\tau_{0}\xi) \nonumber \\
\mbox{} & + & \frac{m_{1}^{2}}{2}
-m_{1}\tanh(m_{1}J) \\
%%%%%%%%%%%%%%%%%%%%%%%
c_{h}\frac{dh}{dt} & = & 
\frac{\sum_{\xi}
{\rm e}^{\beta_{s}m_{0}\xi}}
{2{\cosh}(\beta_{s}m_{0})}
\int_{-\infty}^{\infty}
Dx(\tau x+\tau_{0}\xi){\tanh}
(Jm+h\tau x+h\tau_{0}\xi) -\tau^{2}h
\end{eqnarray}
%%%%%%%%%%%%%%%%
where we should remember that $m$ and $m_{1}$ obey 
the differential equations  
%%%%%%%%%%%%%%%%%%%
%%%%%%%%%%%%%%%%%%%%%
\begin{eqnarray}
\frac{dm}{dt} & = & -m +
\frac{\sum_{\xi}{\rm e}^{\beta_{s}m_{0}\xi}}
{2{\cosh}(\beta_{s}m_{0})}
\int_{-\infty}^{\infty}
Dx \,{\tanh}(Jm+h\tau x+h\tau_{0}\xi) \\
%%%%%%%%%%%%%%%%%%%
\frac{dm_{1}}{dt} & = & 
-m_{1}+{\tanh}(m_{1}J). 
\end{eqnarray}
%%%%%%%%%%%%%%%%
%%%%%%%%%%%%%
By solving these coupled equations, 
we obtain time dependences of the hyper-parameters  
$J(t),h(t)$ and relaxation process of the systems, namely, 
$m(t),m_{1}(t)$. 
In this paper, 
we fix the relaxation times as $c_{J}=c_{h}=1$. 

In FIG. \ref{fig3}, 
we plot time dependences of 
the hyper-parameters $J, h$ and order parameters 
$m, m_{1}$. 
From this figure,   
we see that the final state 
of the hyper-parameters is 
optimal, namely, 
$(J_{*}, h_{*}) \equiv (1/T_s, \beta_{\tau}=\tau_{0}/\tau^{2})=(1.1,1)$ 
and this convergent point is independent of 
the initial conditions. 
%%%%
Time evolutions of the overlap 
$M$ are also plotted in FIG. \ref{fig4} (upper figure). 
We find that 
the overlap $M$ converges to 
the best possible value in FIG. \ref{fig1}.  
In FIG. \ref{fig5}, 
we plot flows of hyper-parameter $J$-$h$. 
From this figure, we find that each flow does not 
take the shortest path to the solution and 
goes a long way around the solution.

%%%%%%%%%%%%%%%%%%%%%%%%%%%%%%%%%%%%%%%%%%%%%%%%%%%%%%%%%%
\subsection{EM algorithm}
%%%%%%%%%%%%%%%%%%%%%%%%%%%%%%%%%%%%%%%%%%%%%%%%%%%%%%%%%%

In the previous subsection, we investigated 
the process of the MML method by gradient decent as 
a dynamics. 
In this section, we analyze the 
performance of {\it EM algorithm} \cite{Dempster} 
as another candidate to maximize the 
marginal likelihood.  

In EM algorithm, we first 
average the log-likelihood 
function 
%%%%%%%%%%%%
\begin{eqnarray}
\log P(\{\tau\}|\{\sigma\})P(\{\sigma\}) & \equiv & 
\frac{J}{N} \sum_{ij}\sigma_{i}\sigma_{j}
+h\sum_{i}\tau_{i}\sigma_{i}
-\log\sum_{\sigma}
\exp \left( 
\frac{J}{N}
\sum_{ij}\sigma_{i}\sigma_{j}
\right) \nonumber \\
\mbox{} & + & \frac{N\tau_{0}}{2\tau^{2}}
-\frac{N\tau^{2}h^{2}}{2}
\end{eqnarray}
%%%%%%%%%%%%%%%%%%%%%
over the time dependent 
posterior distribution 
%%%%%%%%%%%%%%
\begin{eqnarray}
P_{t}(\{\sigma\}|\{\tau\}) & \equiv & 
\frac{{\rm e}^{\frac{J_{t}}{N}\sum_{ij}\sigma_{i}\sigma_{j}
+h_{t}\sum_{i}\tau_{i}\sigma_{i}}}
{\sum_{\sigma}{\rm e}^{\frac{J_{t}}{N}\sum_{ij}\sigma_{i}\sigma_{j}
+h_{t}\sum_{i}\tau_{i}\sigma_{i}}}.  
\end{eqnarray}
%%%%%%%%%%%%%%%%%
This average is referred to as {\it Q-function}. 
As we are interested in the 
averaged behavior of the Q-function, 
we need the following 
averaged Q-function : 
%%%%%%%%%%%%%
\begin{eqnarray}
Q(J,h|J_{t},h_{t}) & \equiv & 
\left[\sum_{\sigma}
P_{t}(\{\sigma\}|\{\tau\})
\log P(\{\tau\}|\{\sigma\})P(\{\sigma\})
\right]_{\{\xi,\tau\}} \nonumber \\
\mbox{} & = & 
J\left[
\frac{\sum_{\sigma}
(\sum_{ij}\sigma_{i}\sigma_{j})
{\rm e}^{\frac{J_{t}}{N}\sum_{ij}\sigma_{i}\sigma_{j}
+h_{t}\sum_{i}\tau_{i}\sigma_{i}}}
{\sum_{\sigma}
{\rm e}^{\frac{J_{t}}{N}\sum_{ij}\sigma_{i}\sigma_{j}
+h_{t}\sum_{i}\tau_{i}\sigma_{i}}}
\right]_{\{\xi,\tau\}} \nonumber \\
\mbox{} & + & h
\left[
\frac{\sum_{\sigma}
(\sum_{i}\tau_{i}\sigma_{i})
{\rm e}^{\frac{J_{t}}{N}\sum_{ij}\sigma_{i}\sigma_{j}
+h_{t}\sum_{i}\tau_{i}\sigma_{i}}}
{\sum_{\sigma}
{\rm e}^{\frac{J_{t}}{N}\sum_{ij}\sigma_{i}\sigma_{j}
+h_{t}\sum_{i}\tau_{i}\sigma_{i}}}
\right]_{\{\xi,\tau\}} \nonumber \\
\mbox{} & - & 
\log \sum_{\sigma}
{\exp}
\left(
\frac{J}{N}\sum_{ij}\sigma_{i}\sigma_{j}
\right)
+\frac{N\tau_{0}^{2}}{2\tau^{2}}
-\frac{N\tau^{2}h^{2}}{2}
\label{Qfunc}
\end{eqnarray}
%%%%%%%%%%%%%%%%%%%%%%%%%%%%%
where we divided the coupling 
constants $J$ and $J_{t}$ by $N$ to 
take a proper thermo-dynamic limit. 

Then, EM algorithm is summarized as follows. 
%%%%
\begin{itemize}
\item
Step 1. \\
Set initial values of the 
hyper-parameters $J_{0}$, $h_{0}$ and $t\leftarrow 0$. 
\item
Step 2. \\
Iterate the following E (expectation) and M 
(maximization) steps until 
an appropriate convergence condition is satisfied. 
%%%%
\begin{itemize}
\item
E step :  Calculate $Q(J,h|J_{t},h_{t})$. 
\item
M step : Update $J_{t}$ and $h_{t}$ by 
\begin{eqnarray*}
J_{t+1} & = & {\arg} \displaystyle{\max_{J}}
Q(J,h|J_{t},h_{t}) \\
h_{t+1} & = & {\arg} \displaystyle{\max_{h}}
Q(J,h|J_{t},h_{t})
\end{eqnarray*}
%%%
and $t \leftarrow t+1$. 
\end{itemize}
\end{itemize}
%%%%%%%%%%%%%%%%%%%%%%%%%%%%%%%%%%%%
%%%%%%%%%%%%%%%%%%%%%%%%%%%%%%%%%%%%%%%%%%%%%%
For our infinite range mean-field MRF model, the 
averages $[\cdots]_{\{\xi,\tau\}}$ in Eq. (\ref{Qfunc}) 
are calculated  by using the replica method and we obtain 
%%%%%%%%%%%%%%%%%%%%%%
\begin{eqnarray}
\frac{Q(J,h|J_{t},h_{t})}{N} & = & 
-\frac{Jm(t)^{2}}{2}
+
\frac{Jm(t) \sum_{\xi}{\rm e}^{\beta_{s}
m_{0}\xi}}
{2{\cosh}(\beta_{s}m_{0})}
\int_{-\infty}^{\infty}
Dx \,{\tanh}(J_{t}m(t)
+h_{t}\tau x+h_{t}\tau_{0}\xi) \nonumber \\
\mbox{} & + & 
\frac{h \sum_{\xi}
{\rm e}^{\beta_{s}m_{0}\xi}
}
{2{\cosh}(\beta_{s}m_{0})}
\int_{-\infty}^{\infty}
Dx (\tau x+\tau_{0} \xi)
{\tanh}(J_{t}m(t)+h_{t}\tau x+h_{t}\tau_{0}\xi) \nonumber \\
\mbox{} & + & \frac{J}{2}m_{1}(t)^{2}
-\log 2 {\cosh}(m_{1}(t)J)
+\frac{\tau_{0}^{2}}{2\tau^{2}}
-\frac{\tau^{2}h^{2}}{2}.
\end{eqnarray}
%%%%%%%%%%%%%%%%%%%%%%%%%%%%%%%%%%%%%%%%%%%
At the next time step, 
$J_{t+1}$ and $h_{t+1}$ are given by 
the conditions 
$\partial Q/\partial J =0$ and 
$\partial Q/\partial h =0$. 
These two conditions lead to non-linear maps :    
%%%%%%%%%%%%%%%%%
\begin{eqnarray}
J_{t+1} & = & 
\frac{1}{m(t)}
{\tanh}^{-1}
{\biggr [}
-\frac{\{m(t)^{2}-m_{1}(t)^{2}\}^{2}}
{2m_{1}(t)} \nonumber \\
\mbox{} & + &  
\frac{m(t)\sum_{\xi}
{\rm e}^{\beta_{s}m_{0}\xi}}
{2m_{1}(t){\cosh}(\beta_{s}m_{0})}
\int_{-\infty}^{\infty}
Dx\, \tanh 
(J_{t}m(t)+h_{t}\tau x +
h_{t}\tau \xi)
{\biggr ]} 
\label{EMdJ} \\
%%%%%%%%%%%%%%%%%%%%%%%%%%%%
h_{t+1} & = & 
\frac{\sum_{\xi}{\rm e}^{\beta_{s}m_{0}\xi}}
{2\tau^{2}{\cosh}(\beta_{s}m_{0})}
\int_{-\infty}^{\infty}
Dx\,(\tau x+\tau_{0}\xi)
{\tanh}(
J_{t}m(t)+h_{t}\tau x+h_{t}\tau_{0}\xi) 
\label{EMdh}.
\end{eqnarray}
%%%%%%%%%%%%%%
In the above non-linear maps, $m(t)$ and $m_{1}(t)$ 
are time dependent 
magnetizations for 
the systems described by the posterior $P(\{\sigma\}|\{\tau\})$ and 
the the prior $P(\{\sigma\})$, respectively. 
%%%%%%%%%%%
By using mean-field treatment,  
we obtain non-linear maps with respect to 
$m(t)$ and $m_{1}(t)$ as 
%%%%%%%%%%%%
\begin{eqnarray}
m(t+1) & = & 
\frac{\sum_{\xi}
{\rm e}^{\beta_{s}m_{0}\xi}}
{2{\cosh}(\beta_{s}m_{0})}
\int_{-\infty}^{\infty}
Dx\, {\tanh}
(J_{t}m(t)+h_{t}\tau x+h_{t}\tau_{0}\xi) 
\label{EMdm} \\
%%%%%%%%%%%%%%%%%%%%%%%%%
m_{1}(t+1) & = & {\tanh}(J_{t}m_{1}(t)) 
\label{EMdm1}.
\end{eqnarray}
%%%%%%%%%%%%%%%%%%%%%%%%%%%%%%%%%%%%%
By solving 
these non-linear 
maps Eqs. (\ref{EMdJ})-(\ref{EMdm1}), 
we obtain the time 
dependence of the 
hyper-parameters 
$J_t, h_t$ and the 
magnetizations $m(t), m_{1}(t)$. 
We plot 
the results in FIG. \ref{fig4} (lower figure), 
FIG. \ref{fig5} and 
FIG. \ref{fig6}. 
From these figures,  
we see that both 
the MML method by gradient descent and 
EM algorithm obtain the 
optimal hyper-parameters 
$(J_{*},h_{*})=(1.1,1)$, however, 
EM algorithm shows faster convergence than 
the MML by gradient descent. 
In addition, 
the flows of EM algorithm in the hyper-parameter space 
are shorter than those of 
the MML by gradient descent. 
From the posterior distribution appearing 
in the Q-function (\ref{Qfunc}), 
we see that performance 
of EM algorithm highly depends on 
the initial choice of the hyper-parameters 
$J_{0}$ and $h_{0}$. 
Therefore, for the systems which 
have lots of local minima, 
the final solution is sensitive to 
the initial condition on the hyper-parameters. 
However, for our model system 
(the infinite range random field Ising model), 
there is no local minima in the marginal likelihood 
function. 
As the result, the final state of 
EM algorithm is independent of the initial 
conditions.

%%%%%%%%%%%%%%%%%%%%%%%%%%%%%%%%%%%%%%%%%%%%%%%%%%%%%%%%%%%%%%%%%%%%%%%%
\section{Summary}
\label{sec5}
%%%%%%%%%%%%%%%%%%%%%%%%%%%%%%%%%%%%%%%%%%%%%%%%%%%%%%%%%%%%%%%%%%%%%%%%

In this paper, we investigated 
dynamical properties of image restoration 
by using statistical mechanics. 
We introduced an infinite range mean-field version of the 
MRF model and 
solved it analytically. 
We derived differential equations with 
respect to the macroscopic order parameters from 
the microscopic Master equation. 
We also studied dynamics of hyper-parameter estimation 
in the context of the maximum marginal likelihood method 
by using gradient descent and EM algorithm.  
For the MML method by gradient descent, 
Boltzmann machine-type learning equations 
were evaluated analytically 
by the replica method. 
On the other hand, 
EM algorithm led to 
non-linear maps and these maps were also evaluated 
analytically. 
We compared 
these two algorithms and found that 
for both algorithms,  
we obtain the 
optimal hyper-parameters. 
We also found that the speed of convergence for 
EM algorithm is 
faster than that of the MML method 
by gradient descent. 
In addition, the paths to the solution in hyper-parameter space 
by EM algorithm are shorter than those of 
the MML by gradient descent. 
Thus, in this paper, 
we could compare two different methods 
to estimate hyper-parameters 
without any computer simulations. 
Our analytical treatments are applicable to studies of 
performance for the other method including 
{\it deterministic annealing EM algorithm}  
\cite{Streit94,UN94}. 
Moreover,  besides image restoration, 
our approach is useful for the 
other problems, for example, 
learning by Bayesian neural networks \cite{Mackay,Frey}, 
time series predictions \cite{Matsumoto2001} or 
density estimation problem \cite{Barkai94}. 
\\
\\

We thank Hidetoshi Nishimori, Masato Okada, Yukito Iba and 
David Saad for fruitful discussions. 
Our special thanks are due to Toshiyuki Tanaka for useful discussions 
and comments.

%%%%%%%%%%%%%%%%%%%%FIG1 
\begin{figure}
\caption{
$1/J (\equiv T_m)$ dependence of the
overlap $M$. 
Temperature of the original 
image is $T_{s}=0.9$ and 
the noise revel is $\beta_{\tau}=\tau_{0}/\tau^{2}=1 
(\tau_{0}=\tau=1)$. 
We set the field $h$ as 
$h=\beta_{\tau}T_s J=(\tau_{0}T_s/\tau^{2})J=0.9J\equiv h_{\rm opt}$ {\sf (a)}
 and $h=1$ {\sf (b)}. 
In the limit of $1/J \rightarrow 0$, we obtain the 
overlap of the MAP estimation. 
In both cases {\sf (a)} and {\sf (b)}, 
the overlap $M$ takes its maximum at $T_{m}=T_{s}=0.9$.
}
\label{fig1}
\end{figure}

%%%%%%%%%%%%%%%%%%%%%%%%%%%FIG2 
\begin{figure}
\caption{ 
$J$-dependence of the 
averaged marginal likelihood $-K$ (upper figure). 
We set $h=0.5,1$ and $h=1.5$. 
We see that $-K$ takes its 
maximum when we choose 
$J,h$ as $J=1.1(=1/T_s)$ and $h=\beta_{\tau}=1$. 
$h$-dependence of the averaged marginal likelihood $-K$ (lower figure). 
We set $J=0.5,1$ and $J=2.1$. 
We see that $-K$ takes its maximum when we choose 
$J,h$ as $J=1.1=(1/T_s)$ and $h=\beta_{\tau}=1$. 
For both figures, 
we chose $(m,m_{1})$ as a solution 
of Eq. (\ref{m}) and $m_{1}=\tanh (Jm_{1})$ for 
$J=1/T_s$ and $h=\beta_{\tau}$. 
}
\label{fig2}
\end{figure}

%%%%%%%%%%%%%%%%%%%%%%%%%%%%%%%FIG3
\begin{figure}
\caption{ 
From the upper left to the lower right, 
time dependences of the 
hyper-parameters $J$, $h$ and the magnetizations $m$, $m_{1}$ 
are plotted.   
In each graph, 
we choose the initial condition 
{\sf (a)} $J(0)=0.45, h(0)=1, m(0)=m_{1}(0)=0.4$, 
{\sf (b)} $J(0)=0.45, h(0)=0.5, m(0)=m_{1}(0)=0.4$, 
{\sf (c)} $J(0)=2.25, h(0)=1, m(0)=m_{1}(0)=0.4$, 
{\sf (d)} $J(0)=2.25, h(0)=0.5, m(0)=m_{1}(0)=0.4$.  
We set true values of the 
hyper-parameters $T_{s}=0.9$, $\beta_{\tau}=1$}
\label{fig3}
\end{figure}

%%%%%%%%%%%%%%%%%%%%%%%%%%%%%FIG4
\begin{figure}
\caption{
Time dependences of the overlap $M$ 
for the case of the MML by gradient descent 
(upper figure) 
and the case of EM algorithm 
(lower figure). 
For both cases, 
we choose the initial condition as  
{\sf (a)} $J(0)=0.45, h(0)=1, m(0)=m_{1}(0)=0.4$, 
{\sf (b)} $J(0)=0.45, h(0)=0.5, m(0)=m_{1}(0)=0.4$, 
{\sf (c)} $J(0)=2.25, h(0)=1, m(0)=m_{1}(0)=0.4$, 
{\sf (d)} $J(0)=2.25, h(0)=0.5, m(0)=m_{1}(0)=0.4$.  
We set true values of the 
hyper-parameters $T_{s}=0.9, \beta_{\tau}=1$. 
We see that for both cases, 
the optimal overlap $M_{\rm opt}$ 
is obtained as a fixed point of the dynamics.}
\label{fig4}
\end{figure}

%%%%%%%%%%%%%%%%%%%%%%%%%%%%%%%%FIG5
\begin{figure}
\caption{Flows in the hyper-parameter space $(J,h)$. 
We set the initial conditions 
$J(0)=J_{0}=0.45, h(0)=h_{0}=1$, and $m(0)=m_{1}(0)=0.4$ 
(upper figure) 
and $J(0)=J_{0}=2.25, h(0)=h_{0}=1$ and $m(0)=m_{1}(0)=0.4$ 
(lower figure). 
True values of the hyper-parameters are  
$J_{*}=1/T_{s}=1.1, h_{*}=\beta_{\tau}=1$. 
For the case of gradient descent (GD), 
the flows go a long way around the 
solution $(J_{*},h_{*})=(1.1,1)$. 
In order to compare the MML by gradient descent 
with the EM algorithm,  
we also plot flows of EM algorithm (EM). 
We see that EM algorithm takes shorter paths than the 
MML by gradient descent.}  
\label{fig5}
\end{figure}

%%%%%%%%%%%%%%%%%%%%%%%%%%%%%%%FIG6
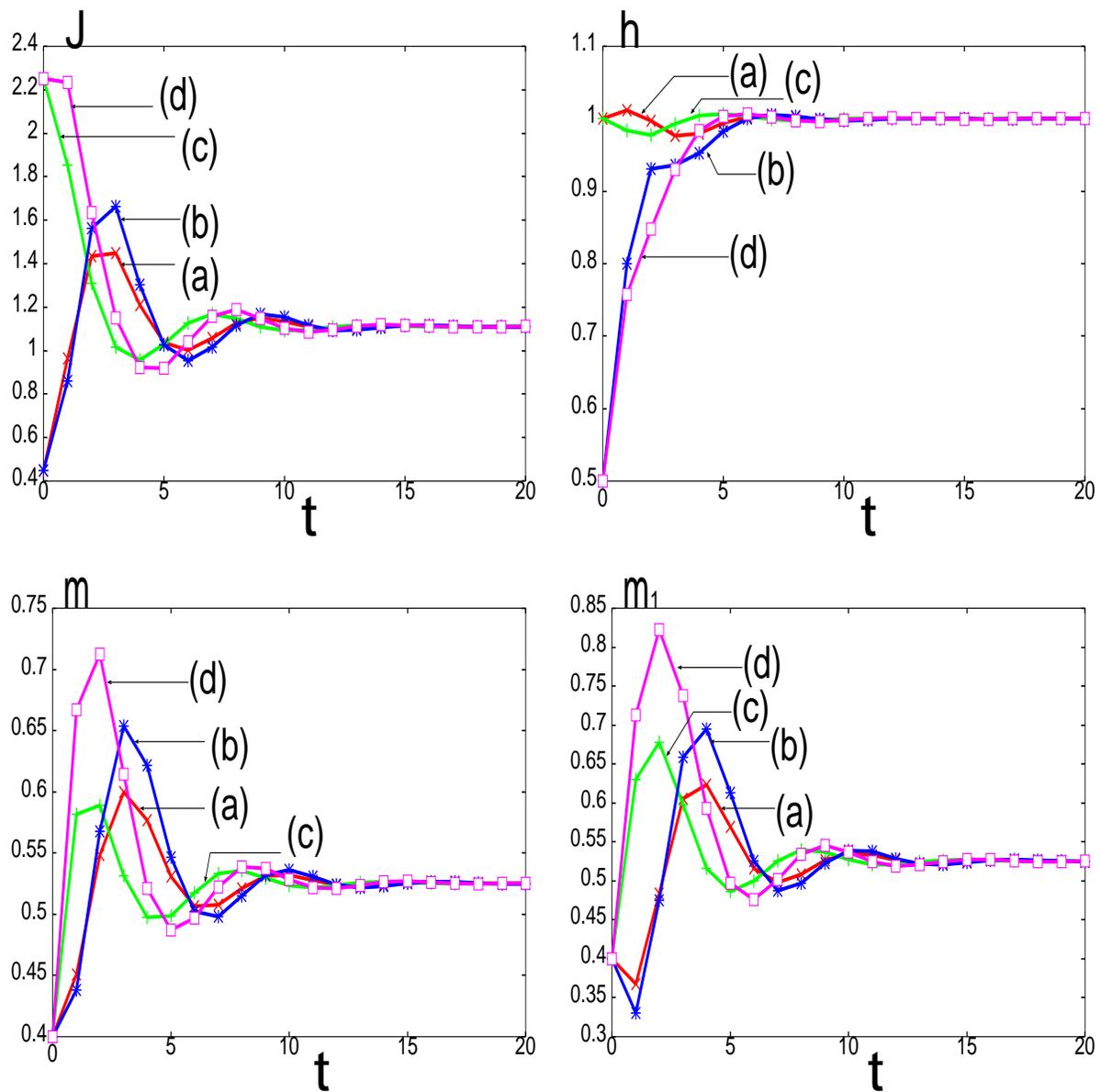
\begin{figure}
\caption{
From the upper left to the lower right, 
time dependences of the 
hyper-parameters $J$, $h$ and the magnetizations $m$, $m_{1}$ 
for the EM algorithm are plotted.   
In each graph, 
we choose the initial condition 
(a) $J_{0}=0.45, h_{0}=1, m(0)=m_{1}(0)=0.4$, 
(b) $J_{0}=0.45, h_{0}=0.5, m(0)=m_{1}(0)=0.4$, 
(c) $J_{0}=2.25, h_{0}=1, m(0)=m_{1}(0)=0.4$, 
(d) $J_{0}=2.25, h_{0}=0.5, m(0)=m_{1}(0)=0.4$.  
We set true values of the 
hyper-parameters $T_{s}=0.9$, $\beta_{\tau}=1$}
\label{fig6}
\end{figure}

%%%%%%%%%%%%%%%%%%%%%%%%%%%%%%%%%%%%%%%%%%%%%%

\end{document}